# The Globalization of Academic Entrepreneurship?

# The Recent Growth (2009-2014) in University Patenting Decomposed

Loet Leydesdorff,*[a] Henry Etzkowitz,[b] and Duncan Kushnir[c]

**Abstract**

The contribution of academia to US patents has become increasingly global. Following a pause, with a relatively flat rate, from 1998 to 2008, the long-term trend of university patenting rising as a share of all patenting has resumed, driven by the internationalization of academic entrepreneurship and the persistence of US university technology transfer. We disaggregate this recent growth in university patenting at the US Patent and Trademark Organization (USPTO) in terms of nations and patent classes. Foreign patenting in the US has almost doubled during the period 2009-2014, mainly due to patenting by universities in Taiwan, Korea, China, and Japan. These nations compete with the US in terms of patent portfolios, whereas most European countries—with the exception of the UK—have more specific portfolios, mainly in the bio-medical fields. In the case of China, Tsinghua University holds 63% of the university patents in USPTO, followed by King Fahd University with 55.2% of the national portfolio.

**Keywords:** university patents, Bayh-Dole Act, nations, IPC, CPC, USPTO

[a] University of Amsterdam, Amsterdam School of Communication Research (ASCoR), PO Box 15793, 1001 NG Amsterdam, The Netherlands; loet@leydesdorff.net ; *corresponding author
[b] Department of Management, Birkbeck, University of London, London WC1E 7HX, UK; h.etzko@googlemail.com
[c] Environmental Systems Analysis, Chalmers University of Technology, Göteborg, Sweden; duncan.kushnir@chalmers.se

# 1. Introduction

University patenting originated in the U.S.A. from the need to protect public health and safety and the university's reputation by controlling the manufacture of drugs and food-related products invented by its staff (e.g., insulin, milk purity analysis devices; Apple, 1989; Bliss, 1982). That income could be generated from licensing patents to manufacturers was an ancillary consequence realized by only a few professors and their universities. Some, like the University of Wisconsin, soon made it a feature of their academic policy, providing a model for later legislation (Etzkowitz, 2015).

The Bayh-Dole Act of 1980 changed the game for university patenting in the US by granting ownership of inventions to universities (and other organizations conducting government-funded research). Prior to the enactment of Bayh-Dole, the US government had accumulated 28,000 patents, but fewer than 5% of these patents were commercially licensed (US General Accounting Office, 1998: 3, at https://en.wikipedia.org/wiki/Bay-Dole_Act ; cf. Berman, 2008; Etzkowitz & Stevens, 1995). The share of patents in the US won by universities grew exponentially for more than two decades (1976-2008). The Bayh-Dole Act was also imitated by other nations as a potential means to bring university research closer to relevant markets (Callaert *et al*., 2013). However, in the decade 1998-2008 university patenting entered a period of relative decline. Leydesdorff & Meyer (2010) discussed this as "the end of the Bayh-Dole effect," while Etzkowitz (2013) warned that the academic analysis of university patenting can suffer from excluding contexts and focusing exclusively on numbers of patents and rates of revenue.

A new growth trend in the share of university patenting is clearly visible since 2008. Which factors are driving this new growth? In recent decades, patents have become more common as an alternative publication outlet for university staff. One can consider university patenting also as a sign of the entrepreneurial transformation of universities (Gibbons *et al*., 1994; Slaughter & Rhoades, 2004); but numbers of patents have not yet been appreciated in major rankings of universities such as the ARWU (Shanghai) or Leiden Rankings.[1] Patenting is expensive,[2] so one can assume that a university, academic scholars, or technology transfer officers must have strong reasons to take the commercial risk of filing for a patent (e.g., Breschi *et al*., 2005; Göktepe-Hulten & Mahagaonkar, 2010; Owen-Smith & Powell, 2001). The reasons for university patenting may extend well beyond financial motives (Etzkowitz and Göktepe, 2015).

The economic effects of academic patents are difficult to specify, but recent efforts suggest a "pebble cast on water effect" of ever-broadening impact on academic research, firm growth, and tax revenues. TTOs often perform a variety of research and regional development functions that may enhance the rate of future applications and also contribute to a penumbra of economic and social development activities. Stevens *et al*. (2016: 139; 143), for example, provide indicative data on firm growth and tax revenues, e.g., 50 billion dollars of the value of the Amgen firm traceable to public sector research, generating 143 billion of private-sector wealth. These authors estimate that five billion in tax impact has derived from 850 million of university royalty income (Swiggart, 2003).

---

[1] The Academic Rankings of World Universities (ARWU) for 2015 can be found at http://www.shanghairanking.com/ ; the Leiden Rankings of top-universities of the Center for Science and Technology Studies at http://www.leidenranking.com/ .
[2] More recently, US law allows a preliminary application to be filed at little cost while commercial potential is explored.

Nonetheless, most universities do not earn from patenting (Geuna & Nesta, 2006). A few universities, like Stanford and NYU, have gained considerably from successful patents. Some universities have lost money by entering this market; others have made huge profits, but typically on a relatively small proportion in a portfolio in which other applications could not succeed at commercialization (Breznitz & Etzkowitz, 2016). Recently (December 9, 2015), Boston University (BU) won a court case about a patent for blue LEDs invented by Theodore Moustakas (USPTO Patent nr. 5,686,738; Nov. 11, 1997). BU was awarded US$13 million for the infringement of this patent by three Taiwan-based companies.

Patents remain indicators of invention, situated at the very beginning of a pipeline that is still far from market introduction and innovation, let alone revenue and profit. The environment can be considered as "hyper-selective" with the odds against newcomers to the market (e.g., Bruckner *et al*., 1994; Dosi, 1982). A plethora of measures have been proposed and implemented—e.g., translational research funds at the university (MIT, Deshpande; UC San Diego, the Von Liebig Centre), at the state level (California Stem Cell Initiative), and at the national level (NIH)—to move the process forward along the innovation process through an "assisted linear model" of innovation, including incubators, accelerators, and regional innovation eco-systems (Etzkowitz, 2006). However, universities often do not patent, especially in incremental engineering topics, but leave the patenting to an industrial partner in compensation for other benefits or ongoing research collaborations.

From an innovation-systems perspective, patenting, and university patenting in particular, can perhaps be considered as early indicators of change. Furthermore, patents at USPTO have been considered as more competitive for emerging markets than patents filed with other national or regional patent offices (Criscuolo, 2004; Jaffe & Trajtenburg, 2002). Note that concepts such as "national innovation systems" (Freeman, 1987; Lundvall, 1988) and "the knowledge-based economy" (David & Foray, 1995) emerged much later in (e.g., OECD) policy documents than the introduction of the Bayh-Dole Act in 1980 (Godin, 2006).

**After "the end of the Bayh-Dole effect"**

After a long period of exponential growth in university patenting in the US (1976-2008), the decade 1998-2008 can be considered as a period of relative decline. Feldman and Clayton (2016) attribute the downturn in 2008 to the global economic recession. However, the period of relative decline antedates the recession, and the recession does not by itself explain the growth since 2008.

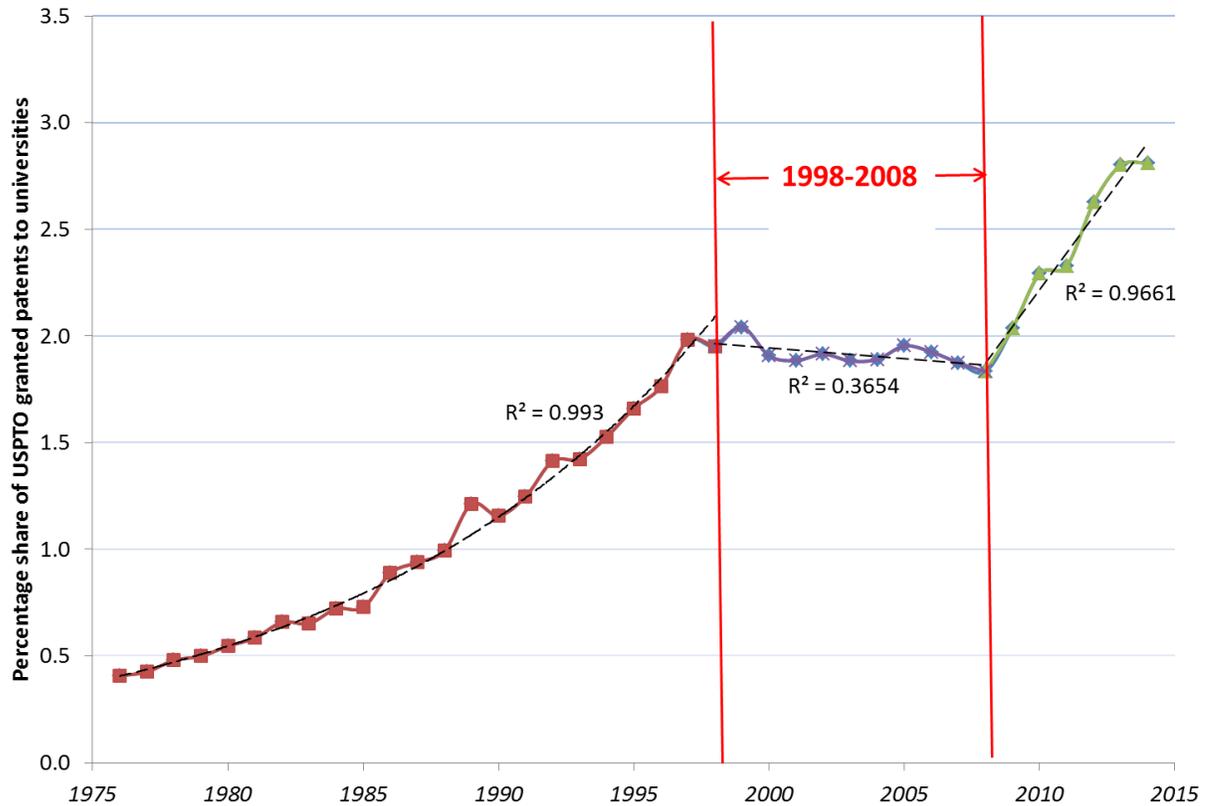

**Figure 1**: Long-term trends of the percentage share of USPTO patents granted to universities and institutes of technology.

Figure 1 analyzes the three periods in terms of their best-fit lines: an exponential upswing until the late 90s, a decline between 1998 and 2008, and resumed linear growth thereafter. Whereas exponential growth in the first period may be indicative for an endogenous self-reinforcing development—presumably triggered by the Bayh-Dole Act (Mowery *et al*., 2001; Sampat, 2006; cf. Kenney & Patton, 2009)—linear growth is more likely the result of an external driver. What may be the independent variables of this upward trend? Leydesdorff & Meyer (2013) suggested that patenting by non-US universities at USPTO could be one of the sources of the upswing.

In order to answer this question in greater detail, we decompose the numbers for the latter period in terms of nations and International Patent Classifications (IPC). International Patent Classifications provide a fine-grained index system of patents worldwide that is now further developed in collaboration between USPTO and the European Patent Organization (EPO) into the system of Cooperative Patent Classifications (CPC).[3] The system is elaborated to the level of 14 digits; but we use the 129 classes at the 3-digit level and the 670 classes at the 4-digit level—that are similar between IPC and CPC—as (however imperfect) indicators of the substantive dimension. In the geographical dimension, the analysis is pursued at the level of nations: which nations are capturing a hold in these high-tech markets by means of university patenting, and in terms of which technologies? Can the patterns inform us about competitive edges? The national portfolios can be decomposed further in terms of lower-level geographical units or specific universities, *mutatis mutandis* (Leydesdorff, Heimeriks, & Rotolo, 2015).

**2. Data and Methods**

USPTO patents are publicly available for download at http://patft.uspto.gov/netahtml/PTO/search-adv.htm . We use online search data and therefore whole-number counting. For the decomposition, USPTO data was additionally batch-downloaded by one of us as a complete set for the period 1976-2014 from Google on October 2, 2015. This set contains 4,965,279 patents ranging from 70,194 patents granted in 1976 to 301,643 in 2014. The analysis is restricted to so-called "utility" or technical patents; design patents and genetic sequences were excluded, and reissued patents are only counted once. This data set is therefore approximately 10% smaller than that obtained by searching online for a given year, with design patents accounting for most of the difference.

---

[3] IPC was replaced with the Cooperative Patent Classification by USPTO and the European Patent Organization (EPO) on January 1, 2013. CPC contains new categories classified under "Y" that span different sections of the IPC in order to indicate new technological developments (Scheu *et al*., 2006; Veefkind *et al*., 2012).

The number of total patents exhibits linear growth during the entire period 1976-2008 ($r^2 > 0.90$) with an increase ($\beta$) of approximately 4,700 patents per year. After 2008, the growth accelerates to more than 25,000 patents granted per year ($r^2 > 0.96$). The increase of university patenting during most of this period is thus part of a general trend, but was reinforced to an exponential trend during the period 1975-1998. The linear trend in university patenting since 2008 is based on an increase of approximately 1,000 patents/year (that is, an increase of 0.16% in the share of USPTO total per year).

For reasons of clarity, we shall express all our findings below as percentages of yearly totals of patents at USPTO and thus normalize for the growth in volume. We use granted patent dates because using filing dates would make our results unreliable for the last few years. The search string used in each consecutive year is 'AN/University OR AN/"Institute of Technology" OR AN/universite OR AN/universitat OR AN/ecole OR AN/universiteit'.[4] The abbreviation "AN" stands for "assignee name" in USPTO. The data for the period 2009-2014 is organized in terms of the 62 nations holding university patents in the database, and both 129 IPC categories at the three-digit level and 670 IPC-4 digit classes. Of these classes, 108 and 385, respectively, were assigned to these patents. IPC classes can also be cross-tabled with the national addresses, so that strength and growth can be indicated for each nation with the different granularities of IPC-3 and

---

[4] The diacritical characters in "école" and "universität" cannot be included online. This search string can be further extended with names in other languages such as "universidad" in Spanish. We return to this issue in the discussion section.

IPC-4. As noted, nations can be decomposed into lower-level units like cities by using, for example, the zip-codes in the address field.

**3. Results**

*3.1. US versus non-US*

Are the recent increases in university patenting due to foreign patenting in the U.S.A.? The Japanese government, for example, heavily subsidizes and rewards patenting by university staff, but in Japan university patenting has nevertheless stagnated at the national level (Nishimura, 2011). Furthermore, one would expect increases of Chinese patenting in the database in recent years, due to the rapid expansion of the Chinese economy and academic entrepreneurship during the period under study.

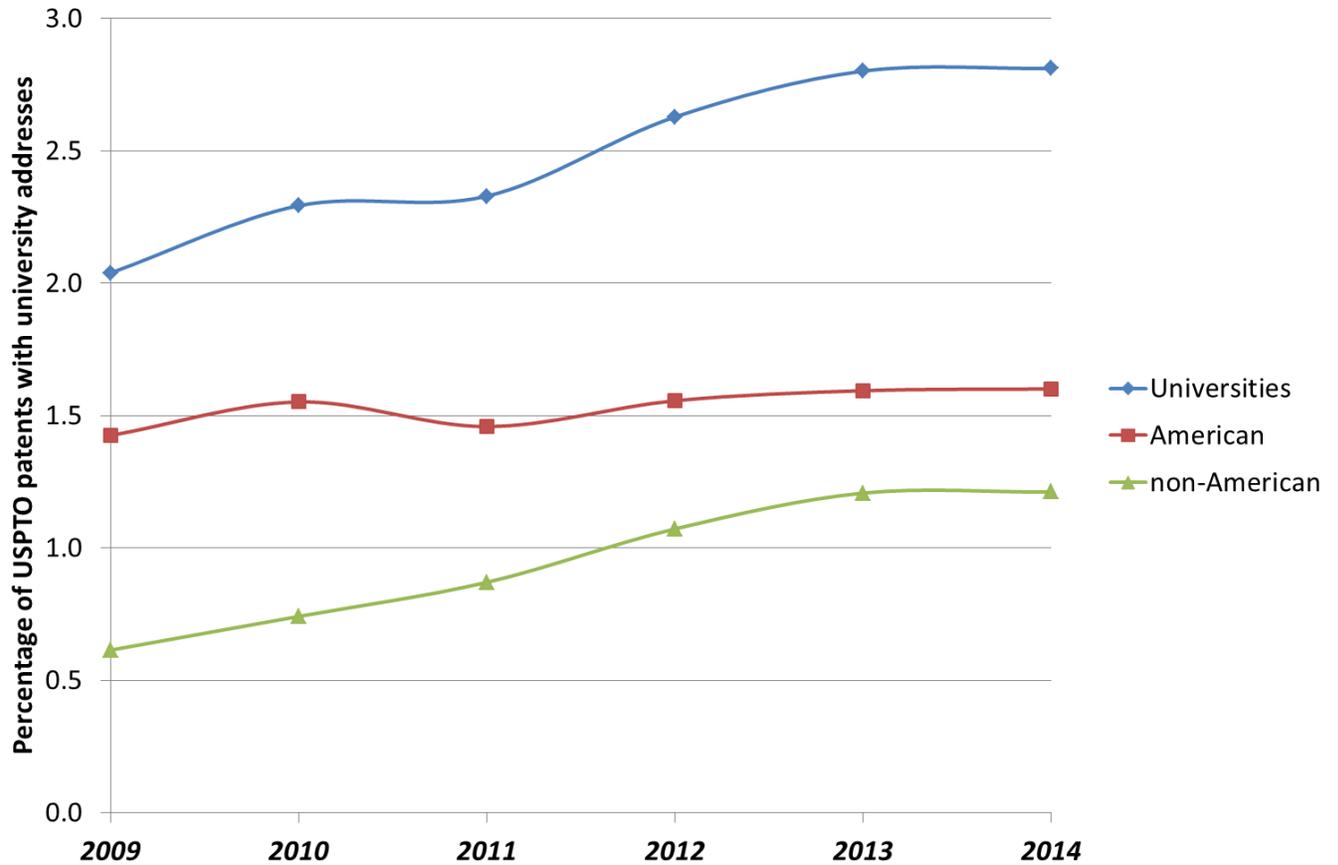

**Figure 2**: US versus non-US university patenting with USPTO. (Data is based on whole-number counting.)

Figure 2 shows the numbers of patents granted to US and non-US universities as percentages of the database. Whereas the numbers tend to stabilize for American universities at an aggregate of almost 1.6%, the proportions of patents granted to non-US universities has doubled during these five years (from 0.6% in 2009 to 1.2% in 2014). (Note that because of the whole-number counting, co-assignments between US and non-US universities are counted as full points in both segments.) As a percentage of the aggregate of patents with university addresses, the American share has declined during these years from 70.1% to 57.5%, while in the overall database the American share is more or less stable (approximately 45%). In sum, the growth is largely due to foreign patenting.

**Table 1**: Countries with growth rates in university patenting larger than the USA; 2009 = 100.

| Country | Volume in 2014 given 2009 = 100 | N of university patents in 2014 |
|---|---|---|
| Saudi Arabia | 1,788 | 143 |
| Norway | 1,300 | 13 |
| India | 1,200 | 48 |
| South Africa | 850 | 17 |
| Korea, Republic of | 459 | 500 |
| Denmark | 457 | 32 |
| Belgium | 429 | 90 |
| China | 381 | 362 |
| Japan | 355 | 720 |
| France | 352 | 236 |
| Taiwan, Province of China | 350 | 888 |
| Ireland | 344 | 31 |
| Israel | 247 | 126 |
| Switzerland | 220 | 55 |
| United Kingdom | 218 | 181 |
| The Netherlands | 207 | 29 |
| Canada | 198 | 192 |
| United States | 191 | 5218 |

Table 1 lists the countries with growth rates in the number of patents granted to universities greater than that of the USA during the period 2009-2014. Although the growth rate of Saudi Arabia is spectacular, the numbers are relatively small, ranging from eight in 2009 to 143 in 2014. The large players and rapid growers, however, are the Asian countries: Taiwan, Korea, Japan, and China. France, Israel, the UK, and Canada are medium-size players, and the other European countries follow with modest contributions (N < 100). India, Norway, and South Africa are rapid growers, but modestly sized. Note that Latin American countries are not on this list. Brazil, for example, holds only 13 university patents granted in 2014; Mexico nine; and Argentina only a single one.[5]

*3.2. Patent classes*

Figure 3 and Table 2 show the decomposition of the growth in terms of 4-digit patent classes assigned to university patents (in USPTO) between 2009 and 2014.

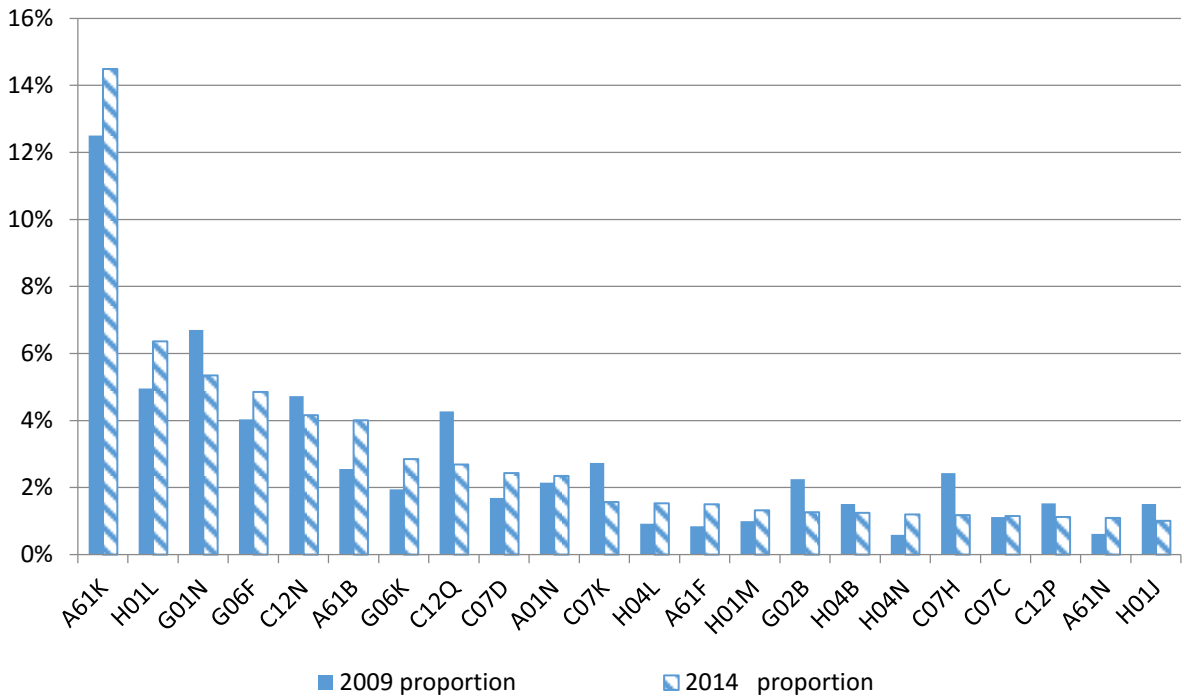

**Figure 3**: Twenty three patent classes that contributed more than 1% to university patenting at USPTO in 2014.

---

[5] These results include the additional terms "universidade" or "universidad" in the online searches.

**Table 2**: Top ten classifications at the 4-digit level of IPC used in university patenting 2014.

| CPC-4 digits | Definition (shortened) | N (2014) | Proportional change 2009-214 |
|---|---|---|---|
| A61K | preparations for medical, dental, or toilet purposes | 1328 | +16% |
| H01L | semiconductor devices; electric solid state devices not otherwise provided for | 583 | +28% |
| G01N | investigating or analysing materials by determining their chemical or physical properties | 490 | -20% |
| G06F | electric digital data processing | 445 | +20% |
| C12N | micro-organisms or enzymes; compositions thereof | 381 | -12% |
| A61B | diagnosis; surgery; identification | 367 | +57% |
| G06K | recognition of data; presentation of data; record carriers; handling record carriers | 261 | +47% |
| C12Q | measuring or testing processes involving enzymes or micro-organisms | 246 | -37% |
| C07D | heterocyclic compounds | 223 | +44% |
| A01N | preservation of bodies of humans or animals or plants or parts thereof | 215 | +9% |

The changes between 2009 and 2014 (in the right-most column of Table 2) are in tens of percentages. Is this indicative of relatively rapid shifts of academic agendas at research fronts?

*3.3. Which nations are increasing their presence in which IPC classes?*

The two main dimensions of the set—the institutional one analyzed here in terms of nations and the substantive one that we try to capture with IPC classes—can also be cross-tabled. This matrix contains a wealth of information:

1. The distributions of IPC classes over nations in the data can be overlaid on Google maps using, for example, the software made available at http://www.leydesdorff.net/software/patentmaps/ (Leydesdorff & Bornmann, 2012; Leydesdorff, Alkemade, Heimeriks, & Hoekstra, 2015).

2. The distributions of nations over IPC classes can be overlaid on the IPC-based maps developed by Leydesdorff, Kushnir, & Rafols (2014), available at http://www.leydesdorff.net/ipcmaps/ .

Figure 4 shows the network of 28 nations versus 69 IPC codes at the three-digit level that forms the ($k = 3$) core group in the 2014 set.[6,7]

---

[6] Eight nodes that are not connected, 44 connected with a single link, and 21 with two links were removed in order to keep the figure readable.
[7] We use the program NetDraw which is particularly suited for visualizing asymmetrical (two-mode) networks (Borgatti, 2002). NetDraw is freely available at https://sites.google.com/site/netdrawsoftware/home.

**Figure 4**: Twenty eight countries and 69 IPC 3-digit categories form the (*k* = 3) core set of university patenting. (USPTO, 2014).

Figure 4 shows that universities in the Asian countries (Japan, China, Korea, and Taiwan) have a pattern of patenting very similar to US universities, whereas European universities (with the exception of Great Britain) share relations to specific patent categories with the U.S.A. The UK assumes an in-between position. (As noted, the number of patents from Saudi Arabia [SA] with a university address is very small.)

**Table 3**: Most frequently present IPC-4 category in national portfolios of university patenting.

| Country | Top category IPC 4-digits 2014 | |
|---|---|---|
| Saudi Arabia | electric digital data processing | G06F |
| Norway | diagnosis; surgery; identification | A61B |
| | preparations for medical, dental, or toilet purposes | A61K |
| India | recognition of data; presentation of data; record carriers; handling record carriers | G06K |
| South Africa | preparations for medical, dental, or toilet purposes | A61K |
| Belgium | preservation of bodies of humans or animals or plants or parts thereof | A01N |
| Korea, Republic of | electric digital data processing | G06F |
| China | semiconductor devices; electric solid state devices not otherwise provided for | H01L |
| Japan | semiconductor devices; electric solid state devices not otherwise provided for | H01L |
| Taiwan, Province of China | semiconductor devices; electric solid state devices not otherwise provided for | H01L |
| Ireland | preparations for medical, dental, or toilet purposes | A61K |
| France | preparations for medical, dental, or toilet purposes | A61K |
| Denmark | processes or means, e.g. batteries, for the direct conversion of chemical energy into electrical energy | H01M |
| Israel | preparations for medical, dental, or toilet purposes | A61K |
| Switzerland | preparations for medical, dental, or toilet purposes | A61K |
| United Kingdom | preparations for medical, dental, or toilet purposes | A61K |
| The Netherlands | micro-organisms or enzymes; compositions thereof | C12N |
| | preparations for medical, dental, or toilet purposes | A61K |
| Canada | preparations for medical, dental, or toilet purposes | A61K |
| United States | preparations for medical, dental, or toilet purposes | A61K |

In Table 3, the IPC-4 classes with relatively the most patents are provided for the same countries as listed in Table 2 above. Most western nations on this list focus on patenting in the bio-medical arena; but Denmark is mainly patenting in energy conversion given its industrial focus on alternative sources of energy. American universities share the focus on the bio-medical category with other Western countries, but as noted above, the *pattern* of patenting at the national level is more akin to that of the four leading Asian nations. The focus is here on electronic devices.

**Table 4**: Leading universities in national portfolios 2014.

| Country | Top-university | N | national share |
|---|---|---|---|
| Saudi Arabia | King Fahd University of Petroleum and Minerals | 79 | 55.2% |
| Norway | Universitetet i Oslo | 5 | 35.7% |
| India | Indian Institute of Technology Bombay | 11 | 22.9% |
| South Africa | University of Cape Town | 5 | 29.4% |
| Korea | Industry-Academic Cooperation Foundation, Yonsei University | 24 | 4.8% |
| Denmark | Technical University of Denmark | 8 | 25.0% |
| Belgium | Universiteit Gent | 12 | 13.3% |
| China | Tsinghua University | 228 | 63.0% |
| Japan | Kyoto University | 37 | 5.1% |
| France | Université Pierre et Marie Curie, Paris 6 | 7 | 3.0% |
| Taiwan | National Tsing Hua University | 113 | 12.7% |
| Ireland | Dublin City University | 8 | 25.8% |
| Israel | Yissum Research Development Company of the Hebrew University of Jerusalem | 11 | 8.7% |
| Switzerland | Ecole Polytechnique Féderale de Lausanne | 18 | 32.7% |
| United Kingdom | University of Birmingham | 10 | 5.5% |
| The Netherlands | Technische Universiteit Delft | 9 | 20.9% |
| Canada | University of British Columbia | 29 | 15.1% |
| United States | Regents of the University of California | 448 | 8.6% |

Table 4 shows the universities leading in these 18 countries in terms of numbers of patents. The patent portfolios are highly skewed in the case of China, where 63% of the university patents are held by Tsinghua University. The King Fahd University follows with 55.2%.

**Conclusions and discussion**

After a decade of relative stagnation (1998-2008), university patenting in USPTO has increased linearly since 2009, rising from approximately 2% to 3% of all annual patents. We have demonstrated that this growth is driven by foreign universities that maintain patent portfolios in theUS The four major players are Taiwan, Korea, China, and Japan, but some smaller players have also begun to patent at USPTO, for example, the King Fahd University in Saudi Arabia. These patents of new entrants and fast growers are mostly concentrated in electronics, whereas a group of moderately growing, mostly European countries patent mainly in the bio-medical sectors.

Our retrievals underestimate the numbers of patents granted to universities a bit, but we do not expect these trends to be different if one adds other possible variants to the search string. The initial extension of the search string from only English words to other languages (French, German, Dutch) did not change the trends significantly. Similarly, the use of online or the batch results are slightly different, with most of the effect from including design patterns or not. But also in this case, the trends expressed in percentages remained robustly the same.

In general, university patenting is just an indicator of output. On the input side, university patenting is driven by contextual factors, including faculty mind-set, university entrepreneurial culture or resistance against that model, research funding levels and other university income, TTO capabilities (in finding licensees and/or encouraging start-ups), and general economic conditions. Patenting is one element in a much broader regime of academic innovation and entrepreneurship (Richards, 2009). As Mitra and Edmonson (2014: 472) formulate: "Patenting represents one way on which universities have become cognizant of their role as exemplary knowledge producers in terms of both public service and the commercialization of such knowledge."

Stevens *et al.* (2016) even argue that "the social act of transferring technology to industry far outweighs the profit earned from such activities." Such a broad socio-innovation framework ("Better World")[11] is now being developed by the Association of University Technology Managers (AUTM) alongside the survey metrics in the Statistics Access for Tech Transfer (STATT) database.[12] Since the economic expectations of academic entrepreneurship remain high in policy discourses, the emerging propensity of non-US universities to patent in the US can be expected to increase further as part of the broader transformation of universities to an entrepreneurial mode in which they play a more significant role in economic and social developments, both on their own initiative and incentivized by national, regional, and multinational actors (OECD, 2012)

---

[11] At http://www.betterworldproject.org/ .
[12] At http://www.autm.net/resources-surveys/research-reports-databases/statt-database-%281%29/

Swiggart, W.(2003) The Bayh-Dole Act & the State of University Technology Transfer in 2003. At www.swiggartagin.com/articles/Bayh_Dole_act.doc, retrieved on Nov. 29, 2015.

Veefkind, V., Hurtado-Albir, J., Angelucci, S., Karachalios, K., & Thumm, N. (2012). A new EPO classification scheme for climate change mitigation technologies. *World Patent Information, 34*(2), 106-111.
22